# Measuring Quasi-Static and Kinetic Coefficient of Restitution Simultaneously using Levitation Mass Method: Experiment and Simulation

Mitra Djamal[a,*], Irfa Aji Prayogi[b], Kazuhide Watanabe[b], Akihiro Takita[b], Yusaku Fujii[b], Sparisoma Viridi[a,c]

[a]Department of Physic, Faculty of Mathematics and Natural Sciences, Institut Teknologi Bandung,

Jalan Ganesa 10, Bandung 40132, Indonesia

[b]School of Science and Technology, Gunma University,

1-5-1 Tenjincho, Kiryu 376-8515, Japan

[c]Physics Department, Technische Universität Dortmund,

Otto-Hahn-Str. 4, D-44221 Dortmund, Germany

*Corresponding Author: mitra@fi.itb.ac.id

## Abstract

Observed peaks in a levitation mass method (LMM), which are usually addressed to signal noise, are discussed in this work. This phenomenon arises due to bounce from plate which is collided by moving part in LMM measurement system. Normally, the moving part and the plate stick very good after their first contact. Simulation using molecular dynamics (MD) method is performed to prove the bounce prognosis.

## Introduction

COR in rocks can be intepreted as ratio betwen work during unloading and loading process upon a sample (Imre *et al*., 2008), which is sligthly different as it kinematicly defined centuries ago (Newton, 1726). Measuring periods of two successive collisions of a bouncing ball on a base is one of the easiest way to obtain COR (Bernstein, 1977), where this method is already well-known for measuring COR of coals (Wooster *et al*., 1945) or by using the Zerbini pendulum with similar concept for rubber (Bassi, 1978). By assumming that the composite COR (Coaplen *et al*., 2004) is dominated only by one object in the case of collision of dissimilier bodies and neglecting other disipassion factors, the loading-unloading COR can be appoximated by kinematic COR.

The quasi-static (loading-unloading) coefficient of restitution (COR) is not exactly the same as the impact (kinematic) COR due to contribution of the bulk of the sample and the duration of contact. The first type of COR is very related to the Young modulus and material dissipation (plastic deformation), while the second type is more to impact coefficient. The bulk of the sample contributes to the first type, while only surface part of the sample contributes to the later. Normally, in a mesurement system only one type of COR can be obtained. But in the system explained in this work, both types of COR can be measured simultanously. The first type is presented in the fom of Young modulus, while the second one is obtained by fit observed bouncing phenomenon with molecular dynamics (MD) simulation result.

Using a levitation mass method (LMM), mass and force measurements for colliding ojects are more accurate (Fujii *et al*., 2010; Fujii *et al*., 2011). The peaks that sometimes are addressed to noise in measureing spring constant of a place can be used to obtain other information, the impact of tested material (kinematic COR).

## Experiment setup

A Copper Beryllium (CuBe) plate is the material under test, which is cut to have such form as shown in Figure 1.

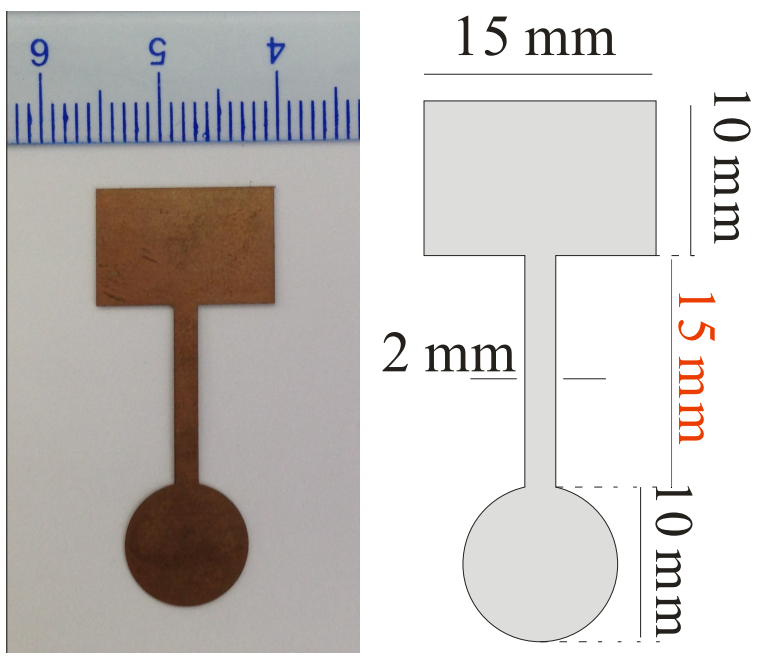

FIGURE 1. CuBe plate with arm length 15 mm, arm width 2 mm, load area diameter 10 mm, thickness 3 mm, total volume 77.56 mm$^3$, density 8.51 × 10$^{-3}$ g/mm$^3$, total masss 0.66 g, and mass (exclude of the rectangle part) 0.277 g.

Rectangle part of the CuBe plate is filrmly clamped at a base of the measurement system, while the round part will be collided by a moving part. The schematic diagram of the measurement system for evaluating dynamic properties of the material is given in Figure 2. In this figure the positive direction for the position, velocity, acceleration, and force acting on the moving part is towards to right.

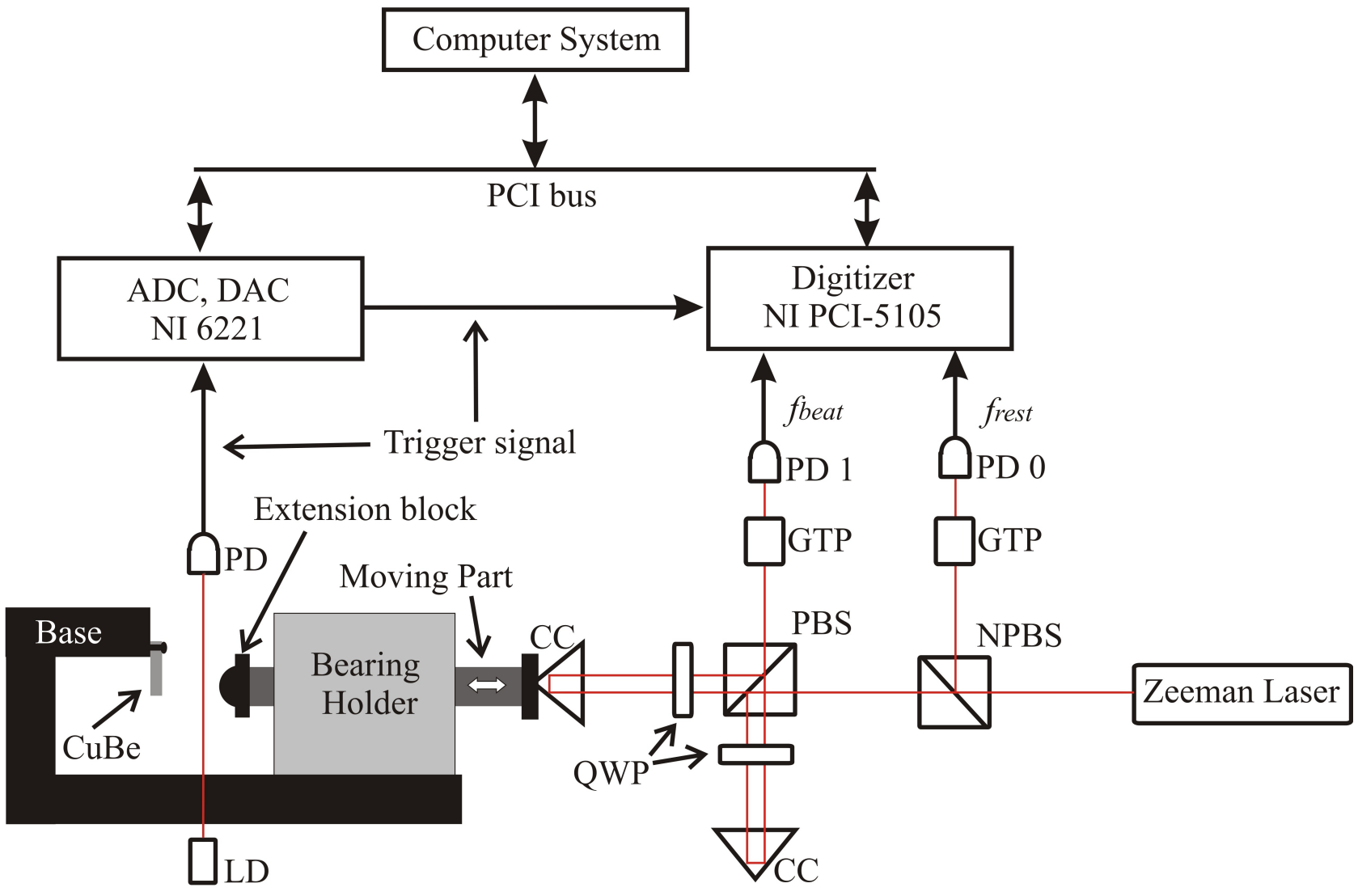

FIGURE. 2. Measurement system for evaluating dynamic properties: PD = Photo Diode, LD = Laser Diode, ADC = Analog-to-Digital Converter, DAC = Digital-to-Analog Converter, CC = Corner Cube, QWP = Quarter Wave Plate, GTP = Gland-Thompson Prism, PBS = Polarizing Beam splitter, and NPBS = Non-Polarizing Beam splitter.

A force is applied to the CuBe plate by smooth hand of moving part at the extension block as it pushing the plate. Then, it would generate a small impact force on the moving part itself as the reaction. The moving part is assembled in such a way inside bearing holder, so that it moves linearly with extremely small friction. The total mass which collides the plate is approximately 17.85 g. At other end of the moving part a corner cube (CC) is attached as seen in Figure 2. The velocities of the moving part when it starts, $v_i$ (m/s), until after impact, $v_f$ (m/s), are accurately measured by interferometer. These velocities depend on the Doppler shift frequency, $f_{\mathrm{Doppler}}$. They can be calculated from

$$v = \frac{\lambda_{\mathrm{air}} f_{\mathrm{Doppler}}}{2} , \quad (1)$$

$$f_{\mathrm{Doppler}} = -\left( f_{\mathrm{beat}} - f_{\mathrm{rest}} \right) , \quad (2)$$

where the $\lambda_{\mathrm{air}}$ is the wavelength of the laser in air (approximately 632.8 nm), $f_{\mathrm{rest}}$ and $f_{\mathrm{beat}}$ are described in the following paragraph.

The light source of interferometer is HP5518A (Zeeman Laser in Figure 2), that generates a pair orthogonally polarizer light beams at two different frequencies. The frequency difference is define as rest frequency, $f_{\mathrm{rest}}$. The laser beam is incident on non-polarizing beam splitter (NPBS). The reflected beam from NPBS is detected by photodetector 0 (PD0), the other is transmitted to the polar beam splitter (PBS). The reflected beam from PBS is incident to first static corner cube (CC), and their polarization directions are rotated by quarter-wave plate (QWP) for $\lambda/4$. It is defined as reference arm. Part of signal beam transmitted from PBS is reflected by second CC which is attached on the moving part, then their frequencies are propagated by Doppler shift, $f_{\mathrm{d}}$. The modulated beams, from the first and second CC are interfered near PBS before detected by photodetector 1 (PD1). The signal of the interference beam is define as beat frequency, $f_{\mathrm{beat}}$. A digitizer, PCI-5105 (National Instument Co., USA), is used to record the intensity waveform of interference signal from both photodetectors at sampling rate 30 Mhz and sampling length 5 Mhz, with measurement duration 0.25 s.

A pneumatic linear bearing, GLS08A50/25-2571 (NSK Co., Ltd., Japan), is attached to an adjustable tilting stage. The maximum weight that supported by the moving part is approximately 1 kg. The stroke of the movement is approximately 25 mm and the thickness of the air film is approximately 10 µm. The tilting angle of the tilting stage was adjusted horizontally with the uncertainty of approximately 0.1 mrad. Measurements in digitizer triggered by a sharp step signal generated using a digital-to-analogue converter. This trigger signal is initiated by a light switch, when the moving part covers the light of laser-diode to the photodiode (PD).

## Simulation

As model for the CuBe plate a cantilever beam is chosen. The Spring constant of an end-loaded cantilever beam for free vibration is given by

$$k_p = \frac{cEt^3 w}{4l^3} , \quad (3)$$

where $E$ is Young modulus, $t$ is the thickness, $w$ is the width, and $l$ is the length. For a rectangular cross section, which is common used in the scanning force microscope (AFM) (Cleveland *et al.*, 1993), value of $c$ is equal to 1 (Stokey, 1989). Equation (3) is modified from its original form using a well-known way (Hoffmann *et al.*, 2000) to accomodate other form though value of $c$. For a certain unique form as shown Figure 1, value of $c$ can be determined empirically by using set of experiments, if it is required.

Collision between moving part and CuBe plate (see Figure 3) is simulated using a soft-sphere scheme, which is common in granular particles collision, implementing only normal force model of linear spring dash-pot (Schäfer *et al.*, 1996)

$$F_N = -k_N \xi - \gamma_N \frac{d\xi}{dt} , \tag{4}$$

where $\xi$ is the overlap between the moving part and the CuBe plate. Numerical integration method used in the simulation is Gear predictor-corrector algorithm of fifth order (Allen *et al.*, 1989), which holds for both Cube plate and moving part motion variabels

$$\begin{bmatrix} x_{0^P}(t+\Delta t) \\ x_{1^P}(t+\Delta t) \\ x_{2^P}(t+\Delta t) \\ x_{3^P}(t+\Delta t) \\ x_{4^P}(t+\Delta t) \\ x_{5^P}(t+\Delta t) \end{bmatrix} = \begin{bmatrix} 1 & 1 & 1 & 1 & 1 & 1 \\ 0 & 1 & 2 & 3 & 4 & 5 \\ 0 & 0 & 1 & 3 & 6 & 10 \\ 0 & 0 & 0 & 1 & 4 & 10 \\ 0 & 0 & 0 & 0 & 1 & 5 \\ 0 & 0 & 0 & 0 & 0 & 1 \end{bmatrix} \begin{bmatrix} x_0(t) \\ x_1(t) \\ x_2(t) \\ x_3(t) \\ x_4(t) \\ x_5(t) \end{bmatrix} , \tag{5}$$

$$\begin{bmatrix} x_0(t+\Delta t) \\ x_1(t+\Delta t) \\ x_2(t+\Delta t) \\ x_3(t+\Delta t) \\ x_4(t+\Delta t) \\ x_5(t+\Delta t) \end{bmatrix} = \begin{bmatrix} x_{0^P}(t+\Delta t) \\ x_{1^P}(t+\Delta t) \\ x_{2^P}(t+\Delta t) \\ x_{3^P}(t+\Delta t) \\ x_{4^P}(t+\Delta t) \\ x_{5^P}(t+\Delta t) \end{bmatrix} + \begin{bmatrix} c_0 \\ c_1 \\ c_2 \\ c_3 \\ c_4 \\ c_5 \end{bmatrix} \Delta x_2 , \tag{6}$$

$$\Delta x_2 = x_2(t+\Delta t) - r_{2^P}(t+\Delta t) , \tag{7}$$

with definition of the non-dimensional motion variabels is

$$x_n(t) = \frac{(\Delta t)^n}{n!} \left( \frac{d^n x_o(t)}{dt^n} \right) . \tag{8}$$

The constants in Equation (6) are 3/16, 251/360, 1, 11/18, 1/6, and 1/60 (Allen *et al.*, 1989). Acceleration obtained from the only interaction between CuBe plate and moving part, Equation (4), is used through Equation (8) for $n = 2$ and then Equations (5) - (7) to produce new motion variabels. Small time step $\Delta t$ must be chosen, e.g. 1/100 or less than the duration time when the Cube plate and moving part are sticked together.

Both, model in Equation (4) and the algorithm in Equations (5) - (8), have been used also for other cases such as modeling of pile stability of a few granular cylinders (Viridi *et al*., 2010), deposition of charged particles (Sustini *et al*., 2011), and ion conduction phenomenon in superionic mateials (Basar *et al*., 2012).

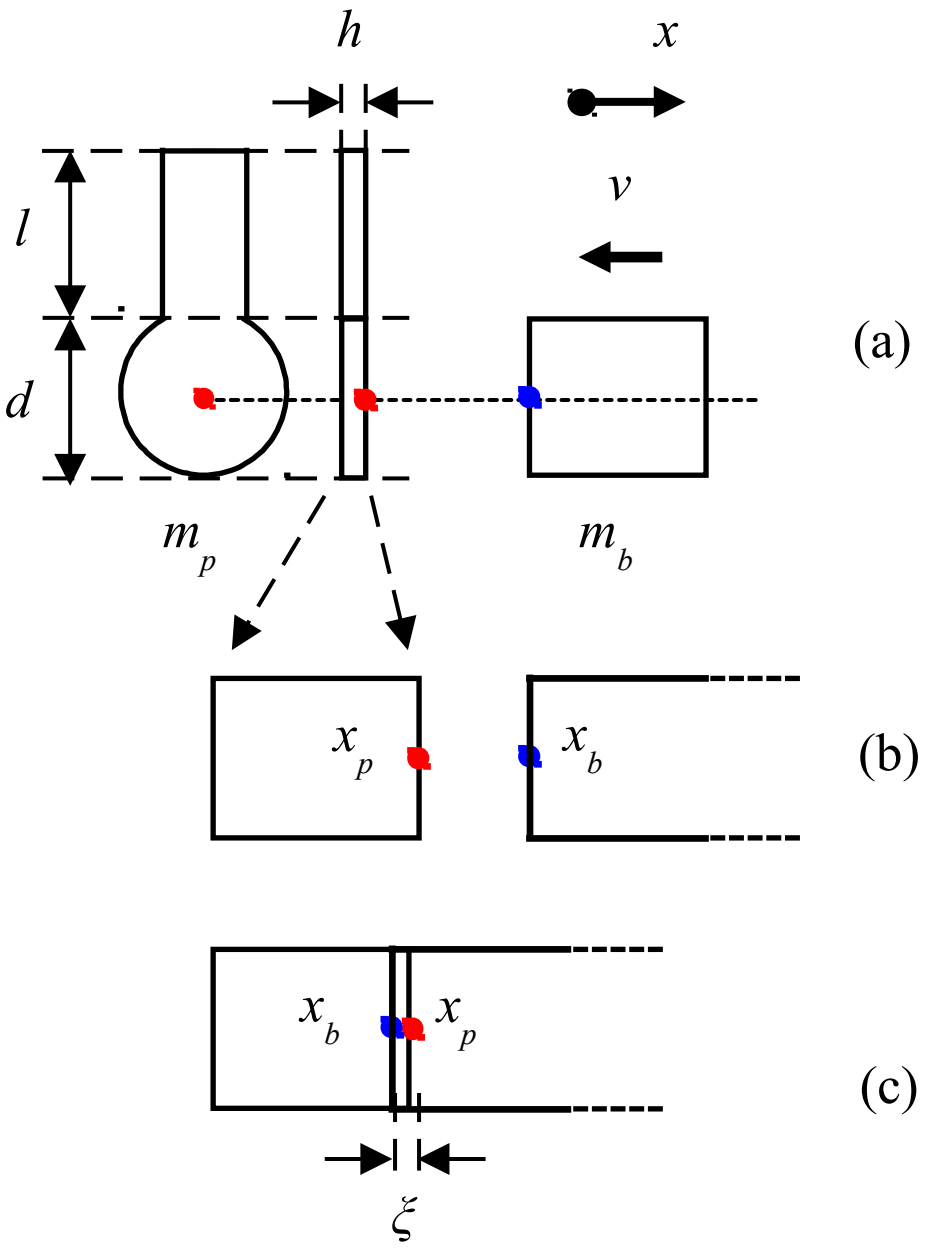

FIGURE 3. Simulation model: (a) CuBe plate and moving block, (b) before collision, no overlap or $\xi = 0$, (c) during collision, overlap $\xi$ exists.

In this model overlap $\xi$ is defined a little bit different than in (Schäfer *et al*., 1996) since the moving block only collides the plate from one direction. It is simply a one dimension collision. Then $\xi$ would be

$$\xi = \max(0, x_p - x_b) \qquad (9)$$

and also its time derivation

$$\dot{\xi} = \max(0, v_p - v_b) \ . \qquad (10)$$

For simplicity, point represents for CuBe plate is taken at the right surface, while for moving part is at the left surface. This choice will leave the need to require thickness of the CuBe plate and length of the moving part.

## Results and discussion

The collision time is represented by the full width half maximum (FWHM) value as shown in Figure 4. It is obtained for the CuBe plate from experiment about 18.09 ms and range of collision time from three different measurements are from 17.24 to 21.36 ms. The force looks like unstable at beginning of collision. It happens since there are some bounces of the CuBe plate while mantaining contact with the moving part. Value of maximum force is observed about 0.5.

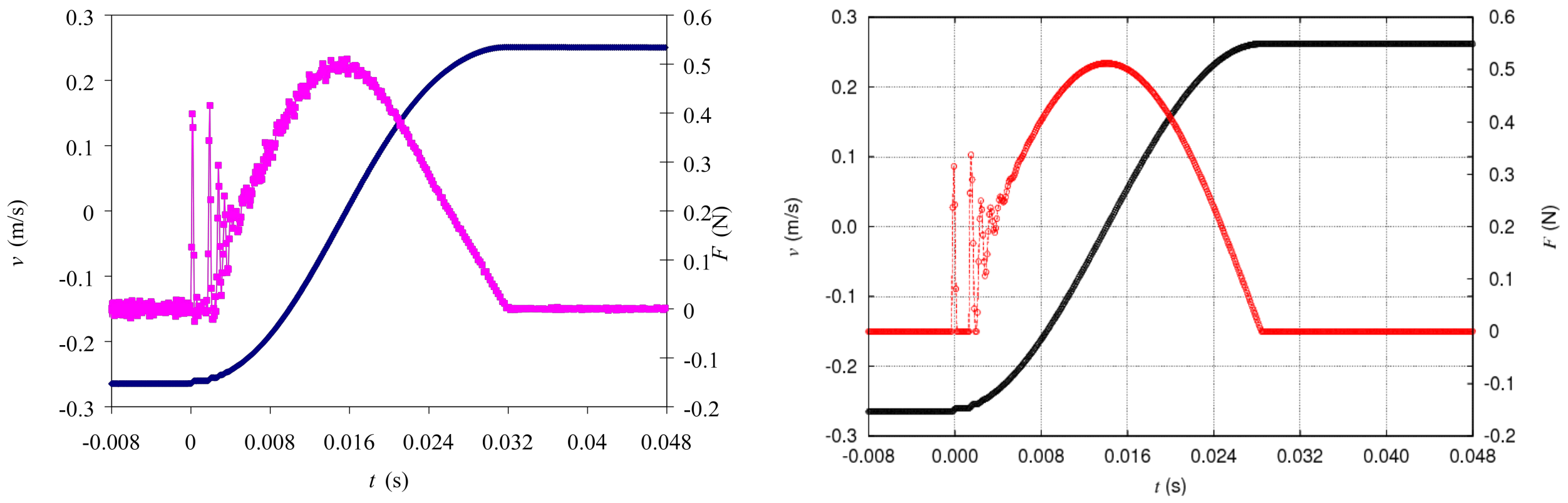

FIGURE 4. Force $F$ and velocity $v$ of moving block as function of time $t$: measured in experiment (left) and calculated from simulation (right).

Following parameters are used in the simulation $m_b = 17.85 \times 10^{-3}$ kg , $v_b = 0.265\ m \cdot s^{-1}$ , $m_p * 0.72\ m_p = 1.984 \times 10^{-4}$ kg (adjusted to tune the simulation results), $k_p = 2.214 \times 10^{2}\ N \cdot m^{-1}$ , $L_p = 2 \times 10^{-2}\ m$ , $k_N = 10^{4}\ N \cdot m^{-1}$ , $\gamma_N = 0.4\ N \cdot s \cdot m^{-1}$ , and $\Delta t = 10^{-7}\ s$ . It is obtained at the end of simulation that $v_b{}' = 0.262\ m \cdot s^{-1}$ . Simulation results in Figure 4 and 5 successfully produce visually similar results to the experiment ones. It means that the parameters value ($m_p$, $k_p$, $k_N$, and $\gamma_N$) are about approximately right.

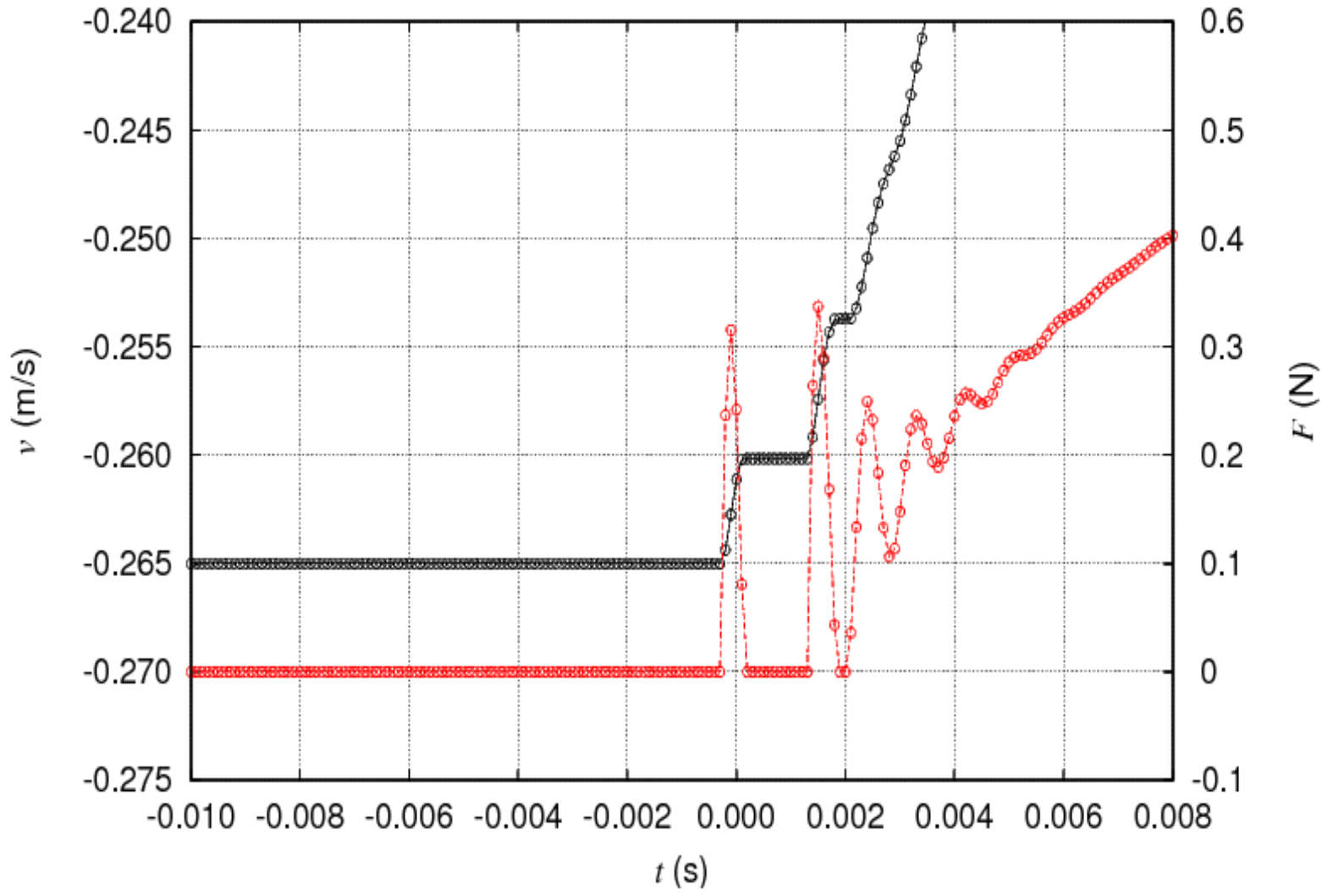

FIGURE 5. Force $F$ and velocity $v$ of moving block as function of time $t$: measured in experiment (left) and calculated from simulation (right), both are zoomed only for certain time around the first collision and bounce phenomena.

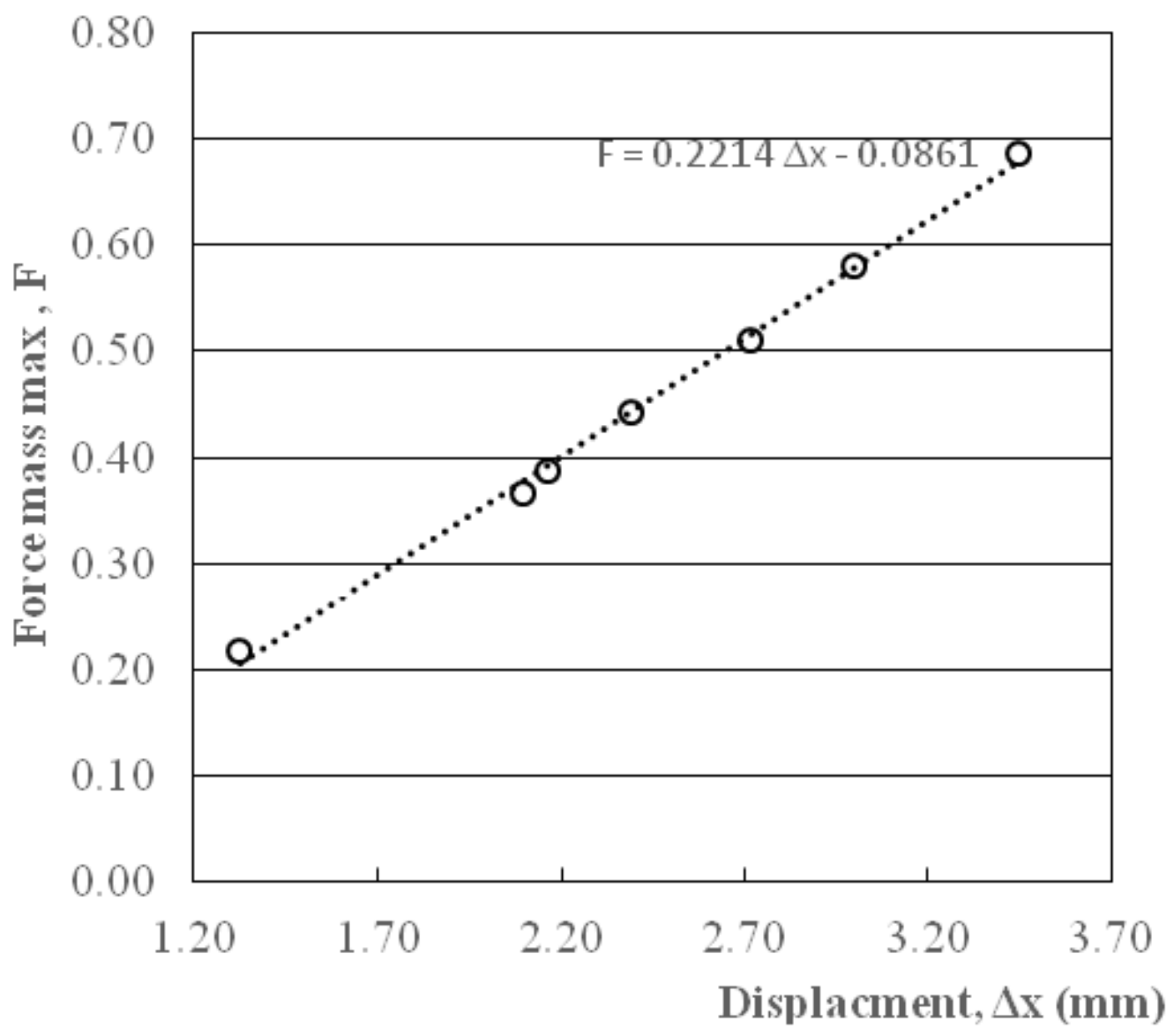

FIGURE 6. Force $F$ against displacement $\Delta x$ to obtain spring constant.

The average spring constant is obtained from slope of linear equation in Figure 6, there uses all data collision from the CuBe plate. The slope expressions by constant value before displacement, $\Delta x$. It shows the Force, $F$ acting on spring plate is proportional to distance (displacement), $\Delta x$ that appropriate with Hooke's law

$$F = k_p \Delta x \quad , \qquad (11)$$

with $k_p$ is spring constant as in Equation (3), which indicates the stiffness of spring. It is obtained that the value of $k_p$ is 0.2214 N/mm. However, there are the other constants, it shows as correction factor for force. It comes since bouncing effect while first contact.

If it is known that Young modulus of CuBe is about 131 GPa (Meaterion Co., 2014; MakeItFrom.com, 2014), then value of $c$ can be calculated from Equation (3), which gives value of 0.01. The reason why this value is quite far from unity, that during the loading-unloading process (while moving block and CuBe plate are sticking together) it seems that mass of the moving block is "added" to the CuBe plate, whereas the mass is about 64 times than the mass of the plate. In the AFM application, where Equation (3) is used, time duration similar to half periode or contact time τ can be calculated from

$$\tau_p = \sqrt{\frac{m_p *}{k_p}} \, , \qquad (12)$$

but in this work it should be

$$\tau_{p+b} = \pi \sqrt{\frac{m_p * + m_b}{k_p}} \quad , \qquad (13)$$

since moving block is attached to the plate nearly all the time (by neglacting the bounces). From the simulation results it is found that $\tau_p \approx 2.97$ ms and $\tau_{p+b} \approx 28.4$ ms. The last prediction is in the same order as to the result from Figure 4, which is about 24 ms.

Spring constant from Equation (3) has an effective mass dependent on the plate geometry. For a rectangular cross section the effective mass $m_p$* is about 0.24 $m_p$ (Cleveland *et al.*, 1993). In this work, simulation results show that it should be 0.72 $m_p$ for producing simulation results similar to experiment results in Figure 4.

Other thing, that could be addressed, is that the round part of CuBe plate could introduced air friction and more inertia (during loading and unloading process) compared to a simple plate with rectangle cross section. This is consistent to the factor 0.72 which is larger than the usual factor 0.24 for effective mass of the plate.

The times where the moving part and CuBe beam collide are given in figure Figure 7. In that figure the blue square shows the difference of $x_p$ and $x_b$ from Equation (5). When it is zero, it means that both object touch each other or collision occurs.

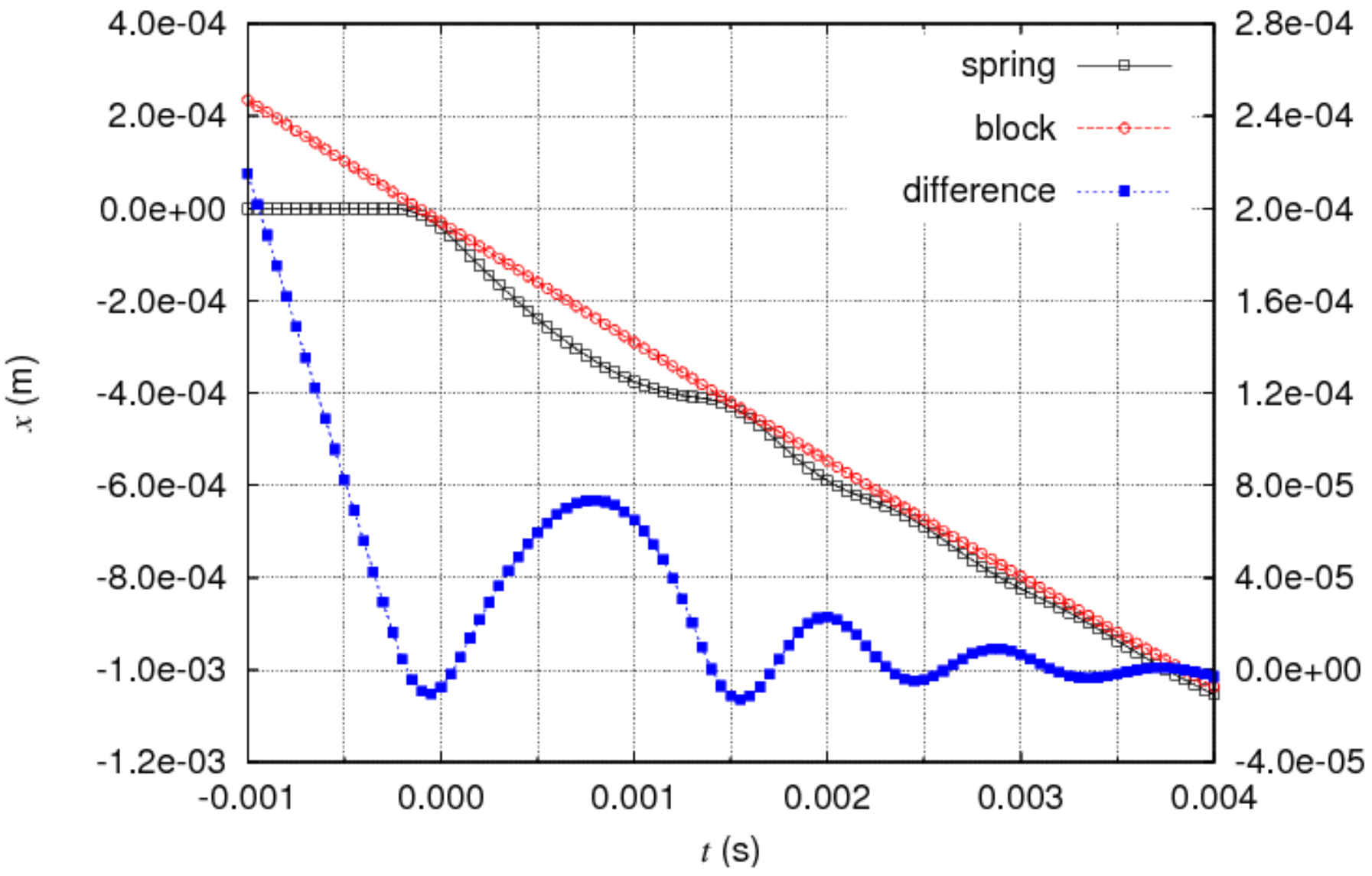

FIGURE 7. Bouncing effect can be seen from position of. Cube plate and moving block. When difference of theses two positions is zero, the two objects collide (blue square).

Quasi-static COR can be obtained from ratio of the second half of parabolic area (loading) in Figure 4 and the first one (unloading), or simply the ratio of reflected and incident velocity from black line in Figure 4, which gives

$COR_{quasi\text{-}static} = |v_r / v_i| = |0.2618907/-0.2650000| = 0.9883$.

Kinetic COR is obtained from Figure 7, which is

$COR_{kinetic} = T_2 / T_1 = 0.00085/0.00150 = 0.5667$.

It has been found that $COR_{quasi\text{-}static} > COR_{kinetic}$, which is contradictory than reported previously (Imre *et al.*, 2008).

## Acknowledgment

Experiment part of this work was supported in part by a research-aid fund of the NSK Foundation for the Advancement of Mechatronics (NSK-FAM) and the Grant-in-Aid for Scientific Research (B) 24360156 (KAKENHI 24360156), while the simulation part is supported by Hibah Desentralisasi DIKTI 2014 and Post-Doc Programme from STRoNG-TiES Erasmus-Mundus 2013.

## Conclusion

It can concluded that two types of COR can be obtained simultaneously by observing the noise-like signal, which sometimes it has not been seen. The fact that $COR_{quasi\text{-}static} > COR_{kinetic}$ is contradictory than reported results by other.

## References

[1] URI http://www.wisetool.com/density.htm (accessed 27 August 2013).

[2] B. Imre, S. Räbsamen, and S. M. Springman, "A coefficient of restitution of rock materials", Computers and Geoscience 34 (4), 339-350 (2008).

[3] I. Newton, "The Principia: Mathematical Principles of Natural Philosophy", 1726, translated by I. Bernard Cohen and Anne Whitman, assisted by Julia Budenz, University of California Press, Berkeley, California, 1999, p. 974.

[4] N. Wooster and W. A. Wooster, "Coefficient of Restitution of Coals", Nature 155 (3948), 787-788 (1945).

[5] A. C. Bassi, "Dynamic Modulus of Rubber by Impact and Rebound Measurements",

[6] A. D. Bernstein, "Listening to the Coefficient of Restitution", American Journal of Physics 45 (1), 41–44 (1977).

[7] J. Coaplen, W. J. Stronge, and B. Ravani, "Work Equivalent Composite Coefficient of Restitution", International Journal of Impact Engineering 30 (6), 581-591 (2004).

[8] J. P. Cleveland, S. Manne, D. Bocek, and P. K. Hansma, "A Nondestructive Method for Determining the Spring Constant of Cantilever for Scanning Force Microscopy", Review of Scientific Instruments 64 (2), 403-405 (1993).

[9] W. F. Stokey, "Shock and Vibration Handbook", McGraw-Hill, New York, 1989, pp. 7.1-7.44.

[10] J. A. Hoffmaan and T. Wertheimer, "Cantilever Beam Vibration", Journal of Sound and Vibration 229 (5), 1269-1276 (2000).

[11] J. Schäfer, S. Dippel, and D. E. Wolf, "Force Scheme in Simulations of Granular Materials", Journal de Physique 1 (6), 5-20 (1996).

[12] M. P. Allen and D. J. Tildesley, "Computer Simulation of Liquids", Oxford University Press, New York, 1989, pp. 82-85

[13] S. Viridi, U. Fauzi, and Adelia, "To Divide or not to Divide: Simulation of Two-Dimensional Stability of Three Grains using Molecular Dynamics", AIP Conference Proceedings 1325 (1), 175-178 (2010).

[14] E. Sustini, S. N. Khotimah, F. Iskandar, and S. Viridi, "Molecular Dynamics Simulation of Smaller Granular Particles Deposition on a Larger One Due to Velocity Sequence Dependent Electrical Charge Distribution", AIP Conference Proceedings 1415 (1), 209-2013 (2011).

[15] K. Basar and S. Viridi, "Simulation of Ion Conduction Phenomenon in Superionic Material using Granular Molecular Dynamics", AIP Conference Proceedings 1454 (1), 215-218 (2012).

[16] "Copper Beryllium Alloys", Materion Co., URI http://materion.com/Products/Alloys/CopperBerylliumAlloys/ (accessed 9 February 2014); "2% Beryllium Copper (Alloy 25, C17200, CuBe2, CW101C)" MakeItFrom.com,

URI http://www.makeitfrom.com/material-data/?for=2-Percent-Beryllium-Copper-Alloy-25-C17200-CuBe2-CW101C (accessed 9 February 2014).

[17] Y. Fujii, T. Jin, and K. Maru, "Precision Mechanical Measurement Using the Levitation Mass Method (LMM)", Indonesian Journal of Physics 22 (1), 1-9 (2011).

[18] Y. Fujii and K. Maru, " Precision Force Measurement Using the Levitation Mass Method (LMM)", Applied Mechanics and Materials 36 (), 41-51 (2010).